\pdfoutput=1
\documentclass[prb,aps,twocolumn,amsmath,amssymb,floatfix,superscriptaddress]{revtex4}
\usepackage{amsmath}
\usepackage{latexsym}
\usepackage{amssymb}
\usepackage{graphics,epstopdf}
\usepackage[colorlinks=true, citecolor=blue, urlcolor=blue ]{hyperref}
\usepackage{epsf,graphics,graphicx}

\date{\today}

\begin{document}
	
\title{Spintronics in antiferromagnetic helix: A new prescription}

\author{Suparna Sarkar}

\email{physics.suparna@gmail.com}

\affiliation {Theoretical Sciences Unit, School of Advanced Materials (SAMat), Jawaharlal Nehru Centre for Advanced Scientific Research, Bangalore 560064, India}

\author{Santanu K. Maiti}

\email{santanu.maiti@isical.ac.in}

\affiliation{Physics and Applied Mathematics Unit, Indian Statistical Institute, 203 Barrackpore Trunk Road, Kolkata-700 108, India}

\begin{abstract}

The occurrence of a finite mismatch between the up and down spin energy channels due to the application of an electric field, leading 
to the generation of a polarized spin current from an unpolarized beam in antiferromagnetic materials, has already been established. 
But, in this work, we report for the first time that even in the absence of any electric field, spin polarization can be achieved. 
We choose a tight-binding antiferromagnetic helix, where the strengths of magnetic moments at different lattice sites are non-uniform. 
The non-uniformity is introduced in two distinct forms, correlated and uncorrelated, and in each case we find a high degree of spin
polarization. The Green's formalism is used to compute the results under various input conditions, and the results are valid 
for a broad range of physical parameters. Our analysis can open up a new direction of getting spin selectivity in different magnetic 
systems with zero net magnetization, in the absence of an electric field.  

\end{abstract}

\maketitle

\section{Introduction}

Spintronics is a highly promising field of research that investigates how to control and employ the spin of electrons~\cite{s1,s2,s3}. 
The groundbreaking discovery of giant magnetoresistance~\cite{fert,grun1,grun2} in the late 1980s opened up new possibilities for 
spin-based electronics. After this experiment, an enormous progress has been made in this area~\cite{sp1,sp2,sp3,sp4} as it offers 
advantages in terms of energy efficiency, memory, data processing speed, storage density and to name a few. With the advancement of 
spintronics, new functionalities like as spin-based transistors, spin-based logic gates, quantum computing and many more are now 
possible to develop in controlled ways~\cite{sp5,sp6,sp7}. 

The primary objective in the field of spintronics is to generate polarized current from an unpolarized beam of electrons using a suitable
functional element. To have this phenomenon, the separation between up and down spin electrons is the primary requirement and that can 
be achieved in presence of different kinds of spin-dependent scattering mechanisms. 
Among the several established proposals~\cite{sp8,sp9,sp10,sp11,sp12,sp13}, one of the common ways is to use the intrinsic properties 
of the materials such as, spin-orbit (SO) coupling~\cite{so1,so2,so3,so4,so5}. In condensed matter systems, Rashba~\cite{rashba}
and Dresselhaus~\cite{dressel} SO couplings are the two commonly used SO couplings and they correspond to the interaction between 
spin and orbital motion of electrons. The Dresselhaus effect arises due to the absence of inversion symmetry in the bulk of a system
and therefore its control by external mechanism is no longer possible. On the other hand, the Rashba SO coupling is generated due to the
breaking of symmetry in confining potential, and it can be tuned externally~\cite{ex1,ex2,ex3,ex4}.
Using this intrinsic property, people have studied spin-dependent phenomena considering different tailor made geometries, molecular
systems~\cite{pr1,pr2,pr3,pr4,pr5,pr6}. But the major problem is that, the SO coupling strength is usually very weak~\cite{smSO},
and hence, a large mismatch among up and down spin channels is no longer possible which practically hinders to achieve a favorable 
spin polarization. One possible route to overcome this issue is to use of ferromagnetic materials, as they have strong spin-dependent
scattering strength~\cite{fm1,fm2} which naturally shifts up and down spin energy channels in a large extent. But, there are some important
restrictions. For instance, achieving of controlled spin-dependent transport phenomena with the help of external magnetic field is a 
challenging task, as the confining of magnetic field in a small scale region is itself too difficult. Moreover, the functionality 
of ferromagnetic materials is often limited to a specific temperature range. All these issues can be circumvented by replacing SO coupled 
and ferromagnetic systems by antiferromagnetic (AF) ones. The antiferromagnetic materials become promising candidates for spin based 
devices due to their unique properties and advantages~\cite{afm1,afm2,afm3}. Unlike ferromagnetic materials, AF systems possess 
inherent stability and robustness due to their balanced magnetic moments and absence of stray fields~\cite{adafm1,adafm2}. In addition, 
they can be operated in the very high frequency range.  

Among many, helical geometries have been the objects of intense research in the field of spintronics due to their unique and diverse
characteristic features~\cite{qfs1,qfs2,qfs3,ss}. The interest has rapidly been picked up following the pioneering work of G\"{o}hler 
{\em et al.}~\cite{ghl} where they have experimentally shown that self-assembled monolayers of DNA molecules can produce highly polarized 
spin current and that even be observed in the limit of room temperature. The central physics is involved with the chirality of the 
molecular system. Changing the handedness, left-handed to right-handed and vice versa, electrons of one specific spin can be transferred 
through the molecule, while the other spin electrons can substantially be suppressed. They refereed to this phenomenon as {\em chiral 
induced spin selectivity} (CISS) effect. Later, many researchers have actively involved in doing research considering different kinds 
of helical geometries, and it has already been established that helicity has an important role in having much superior performance than 
the helicity-free geometries~\cite{ciss1,ciss2,ciss3,ciss4,ciss5}. Due to the several advantages of helical geometries and 
antiferromagnetic materials, people have studied the spin-dependent transport considering antiferromagnetic 
helices~\cite{afmspin1,afmspin2}. The common perception suggests that an antiferromagnetic system cannot produce spin polarized current 
since its net magnetization is zero, and thus the up and down spin Hamiltonians are symmetric to each other. But that symmetry can be 
broken in a helical geometry by applying an electric field, and this fact has already been established by some groups including 
us~\cite{skm1,qfs2}.

In the present work, we put forward a new prescription where spin polarization through an antiferromagnetic helix (AFH) can be achieved even 
in the {\em absence} of any electric field. This is the primary motivation of our work. The other important motivation is that, depending 
on the conformation of the system, we can get long-range or short-range hopping of electrons. Thus, a physical system beyond usual
nearest-neighbor hopping (NNH) can be studied, and it is well known that higher order electron hopping always provides some interesting 
features than the usual NNH model. Motivated with all these facts, we consider an antiferromagnetic helical system as a functional element 
for spin polarization. As already pointed out, {\em the central focus of our work is to establish a polarized spin current from an 
unpolarized beam in the absence of any electric field}. The physical system is described within a tight-binding (TB) framework. Each site 
of the helix contains a finite magnetic moment, and the magnetic moments at different lattice sites are arranged in such a way that the 
net magnetization becomes zero. We choose the magnetic moments non-uniformly in some specific forms which breaks the symmetry of the 
system, and similar kinds of arrangements have also been reported earlier~\cite{nonuniform1,nonuniform2}, which gives us the confidence 
that our proposed systems can also be realized. All the results are worked out based on Green's function formalism. Quite interestingly 
we find that, a high degree of spin polarization can be obtained under different input conditions and the response is valid for a broad 
range of physical parameters.    
 
The outline of the work is as follows. Following the above brief introduction (Sec. I), in Sec. II we describe the helical geometry and 
the required mathematical steps for calculating the numerical results. All the findings are presented and thoroughly discussed in Sec. III.
Finally, the summary of the work is given in Sec. IV. 

\section{Quantum system, tight-binding Hamiltonian, and theoretical formulation}

\begin{figure}[ht]
{\centering \resizebox*{8.5cm}{8.5cm}{\includegraphics{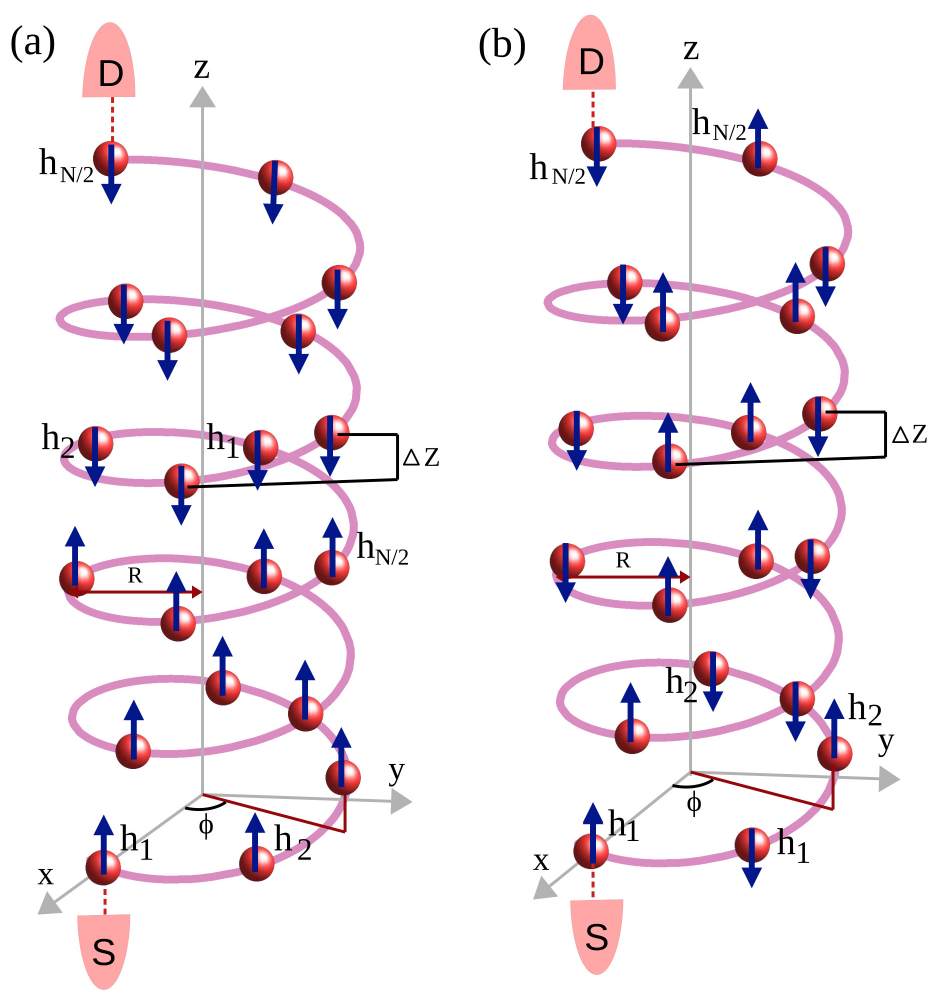}}\par}
\caption{(Color online). Two different spin polarization setups are shown considering a single stranded antiferromagnetic helix coupled 
to source and drain. For setup-1, the first $N/2$ sites exhibit magnetic moments pointing in the $+Z$ direction, whereas the remaining 
$N/2$ sites have magnetic alignment in the opposing direction (Fig.~\ref{model}(a)), and, for setup-2, the adjacent magnetic moments are
arranged in an antiparallel configuration along $\pm Z$ directions (Fig.~\ref{model}(b)). Here, $R$ denotes the radius, $\Delta \phi$ 
represents the twisting angle, and $\Delta z$ corresponds to the stacking distance between the neighboring lattice sites.}
\label{model}
\end{figure}

\subsection{Model and the Hamiltonian}

Let us start with spin polarized setup given in Fig.~\ref{model}, where in each case an antiferromagnetic helix is coupled to two 
contact electrodes, source (S) and drain (D). When an itinerant electron interacts with local magnetic moments of the helix, 
spin-dependent scattering occurs. If $\langle \overrightarrow{S_i} \rangle$ is the average spin at any site $n$ (say), then the 
spin-moment scattering is described as $J \langle \overrightarrow{S_i} \rangle.\vec{\sigma}$, where $J$ is the coupling strength 
and $\vec{\sigma}$ is the Pauli spin vector. This can be written in a compact way as $\vec{h}_i.\vec{\sigma}$, where 
$\vec{h}_i=J \langle \overrightarrow{S_i} \rangle$. We define $\vec{h}_i$ as the spin-dependent scattering parameter.
Two different arrangements of local spins are considered in two distinct setups that are presented in Figs.~\ref{model}(a)
and (b). In one case (Fig.~\ref{model}(a)), the magnetic moments in the first $N/2$ sites are oriented along $+Z$ direction, and 
for the other $N/2$ sites they are aligned in the opposite direction. While, in the other setup (Fig.~\ref{model}(b)), the 
neighboring magnetic moments are oriented in the antiparallel configuration $(\pm Z)$. 
The helical system is defined geometrically by its radius $R$, stacking distance $\Delta z$, and twisting angle $\Delta \phi$. Two distinct
types of helical structures can be identified depending on the structural parameters: the long-range hopping (LRH) helix and the 
short-range hopping (SRH) helix. In the case of LRH, the atoms are positioned closely together, allowing electrons to hop between all 
possible sites. On the other hand, SRH involves atoms that are relatively far apart, resulting in electron hopping being limited to only 
a few neighboring sites. In this work, we consider both these SRH and LRH antiferromagnetic helices, and investigate how each of these 
helical geometries affects the spin-selective transmission.

In each spin polarized setup, $h_i$'s are taken non-uniformly. Two different non-uniform distributions are used those are correlated 
(deterministic) and uncorrelated (random). For the correlated case we choose $h_i$'s following the relation
\begin{eqnarray}
h_i=h_0 |\cos (2\pi b i)| \hskip 1cm  1\le i \le \frac{N}{2}
\end{eqnarray}
where $h_0$ is the cosine modulation strength, $i$ is the site index, and $b\, (=(\sqrt{5}-1)/2)$ is an irrational number
that introduces aperiodicity in the magnetic moments~\cite{gm}.
On the other hand, for the uncorrelated case, $h_i$'s are taken randomly from a ``Box" distribution function of width $h_0$.

For both the setups, the general Hamiltonian of the full system (source-helix-drain) can be written as
\begin{eqnarray}
H & = & H_{\mbox{\tiny helix}} + H_{\mbox{\tiny S}} +  H_{\mbox{\tiny D}} + H_{\mbox{\tiny cpl}}
\label{equ1}
\end{eqnarray}
where $H_{\mbox{\tiny helix}}$, $H_{\mbox{\tiny S}}$, $H_{\mbox{\tiny D}}$, and $H_{\mbox{\tiny cpl}}$ represent the Hamiltonians of 
the helical molecule, source, drain, and the coupling between the helical molecule and the electrodes, respectively. The specific forms
of these Hamiltonians are as follows.

The TB Hamiltonian of the helical molecule with higher order hopping reads as
\begin{eqnarray}
	H_{\mbox{\tiny helix}} & = & \sum_i \mbox{\boldmath{$c$}}_i^{\dagger} 
	\left(\mbox{\boldmath{$\epsilon$}}_i-\mbox{\boldmath{$h$}}_i.\mbox{\boldmath{$\sigma$}}\right)\mbox{\boldmath{$c$}}_i
	\nonumber \\
	& + & 
	\sum_{i=1}^{N-1}\sum_{j=1}^{N-i}\left(\mbox{\boldmath{$c$}}_i^{\dagger}\mbox{\boldmath{$t$}}_j \mbox{\boldmath{$c$}}_{i+j} + 
	\mbox{\boldmath{$c$}}_{i+j}^{\dagger}\mbox{\boldmath{$t$}}_j^{\dagger}\mbox{\boldmath{$c$}}_{i}\right)
	\label{equ2}
\end{eqnarray} 
where different matrices of Eq.~\ref{equ2} are given by
$\mbox{\boldmath{$\epsilon$}}_i=\begin{pmatrix}
	\epsilon_i & 0\\
	0 & \epsilon_i
\end{pmatrix}$,
$\mbox{\boldmath{$h$}}_i.\mbox{\boldmath{$\sigma$}}=h_i\begin{pmatrix}
	\cos\theta_i & \sin\theta_i e^{-k\varphi_i}\\
	\sin\theta_i e^{k\varphi_i} & -\cos\theta_i
\end{pmatrix}$,
$\mbox{\boldmath{$c$}}_i=\begin{pmatrix}
	c_{i\uparrow} \\
	c_{i\downarrow} 
\end{pmatrix}$,
$\mbox{\boldmath{$c$}}_i^{\dagger}=\begin{pmatrix}
	c_{i\uparrow}^{\dagger} & c_{i\downarrow}^{\dagger} 
\end{pmatrix}$,
and
$\mbox{\boldmath{$t$}}_j=\begin{pmatrix}
	t_j & 0\\
	0 & t_j
\end{pmatrix}$. 
Here $\epsilon_i$ is the on-site energy of an electron at site $i$, in the absence of any magnetic interaction. 
$c_{i\sigma}^{\dagger}$ ($c_{i\sigma}$) represents the fermionic creation (annihilation) operator at $i$th site with spin 
$\sigma$ ($\uparrow, \downarrow$). In spherical polar coordinate system, the general orientation of $\overrightarrow{h_i}$ is described 
by the polar angle $\theta_i$ and azimuthal angle $\varphi_i$. The value of $k$ is $\sqrt{-1}$. $t_j$ represents the hopping between the 
sites $i$ and ($i+j$). The expression of $t_j$ looks like~\cite{qfs2}
\begin{eqnarray}
t_j=t_1 e^{-(l_j-l_1)/l_c}
\label{equ3}
\end{eqnarray}
where $t_1$ and $l_1$ are the nearest-neighbor hopping strength and the distance, respectively. $l_c$ is the decay constant, and it has a
dominant role in the electronic hopping strength. $l_j$ measures the distance between the sites $i$ and $i+j$. In terms of the structural
parameters of the helical molecule, the expression of $l_j$ reads as~\cite{qfs2}
\begin{eqnarray}
l_j=\sqrt{[2R\sin(j\Delta\phi/2)]^2+(j\Delta z)^2}.
\label{equ4}
\end{eqnarray}

Now, we describe the TB Hamiltonians of the source, drain and the coupling of electrodes with the helical molecule. The electrodes are 
assumed to be  one-dimensional (1D), perfect, and non-magnetic in nature. The assumption of using strictly one-dimensional electrodes 
does not hamper the underlying physics, as we are working within the wide-band limit approximation, where the allowed bandwidth of the 
helical system is smaller than the bandwidths of the contact electrodes. With nearest-neighbor hopping parameter $t_0$ and 
on-site energy $\epsilon_0$, the sub-Hamiltonians $H_{\mbox{\tiny S}}$ and $H_{\mbox{\tiny D}}$ can be expressed as
\begin{equation}
	H_{\mbox{\tiny S}} =\sum_{n<1}\mbox{\boldmath{$a$}}_n^{\dagger} 
	\mbox{\boldmath{$\epsilon$}}_0\mbox{\boldmath{$a$}}_n +
	\sum_{n<1} (\mbox{\boldmath{$a$}}_n^{\dagger} \mbox{\boldmath{$t$}}_0 
	\mbox{\boldmath{$a$}}_{n-1}+h.c.)
\end{equation}
and
\begin{equation}
	H_{\mbox{\tiny D}} = \sum_{n>N} \mbox{\boldmath{$b$}}_n^{\dagger} 
	\mbox{\boldmath{$\epsilon$}}_0 \mbox{\boldmath{$b$}}_n +
	\sum_{n>N} (\mbox{\boldmath{$b$}}_n^{\dagger} \mbox{\boldmath{$t$}}_0 
	\mbox{\boldmath{$b$}}_{n+1}+h.c.)
\end{equation}
where $\mbox{\boldmath{$\epsilon$}}_{0}=\begin{pmatrix}
	\epsilon_0 & 0\\
	0 & \epsilon_0
\end{pmatrix}$
and
$\mbox{\boldmath{$t$}}_{0}=\begin{pmatrix}
	t_{0} & 0\\
	0 & t_{0}
\end{pmatrix}$.

\vskip 0.2cm
\noindent
The operators $a_n^\dagger, a_n (b_n^\dagger, b_n)$ are the creation and annihilation
operators for source (drain) respectively.

Finally, we write the coupling Hamiltonian $H_{\mbox{\tiny cpl}}$ as
\begin{equation}
	H_{\mbox{\tiny cpl}} = \mbox{\boldmath{$a$}}_0^\dagger
	\mbox{\boldmath{$t$}}_S \mbox{\boldmath{$c$}}_1 +
	\mbox{\boldmath{$c$}}_N^\dagger \mbox{\boldmath{$t$}}_D
	\mbox{\boldmath{$b$}}_{N+1}+h.c.
\end{equation}
where $\mbox{\boldmath{$t$}}_{S}=\begin{pmatrix}
	t_S & 0\\
	0 & t_S
\end{pmatrix}$
and
$\mbox{\boldmath{$t$}}_{D}=\begin{pmatrix}
	t_{D} & 0\\
	0 & t_{D}
\end{pmatrix}$.

\vskip 0.2cm
\noindent
The diagonal elements $t_S$ and $t_D$ represent the coupling strengths of the helical molecule with the source and drain respectively.

\subsection{Theoretical prescription}

The central focus of this study is to inspect the behavior of spin polarization coefficient ($P$) under different input conditions.
We define spin polarization coefficient as~\cite{spl1}
\begin{equation}
	P=\bigg|\frac{I_{\text{sp}}}{I_{\text{ch}}}\bigg| \times 100\%.
	\label{equ8}
\end{equation}
Here, $I_{\text{sp}}=I_\uparrow-I_\downarrow$ and $I_{\text{ch}}=I_\uparrow+I_\downarrow$ are the spin and charge currents, respectively.
Our aim will be to find a high degree of spin polarization as much as we can achieve in the feasible parameter range. The spin polarization
reaches maximum ($P=100 \%$) when only up or down spin electrons transmit, and it become zero when both the spin electrons propagate 
equally. For the other cases, intermediate values are found. 

$I_{\sigma}(\sigma=\uparrow,\downarrow)$ is the spin-specific junction current, where
$I_{\uparrow}=I_{\uparrow\uparrow}+I_{\downarrow\uparrow}$
and $I_{\downarrow}=I_{\downarrow\downarrow}+I_{\uparrow\downarrow}$. We compute the currents following
the Landauer-B\"{u}ttiker formalism through the relation~\cite{gf1,gf2}
\begin{equation}
	I_{\sigma\sigma^\prime} = \displaystyle \frac{e}{h} \int 
	\limits_{E_F-\frac{eV}{2}}^{E_F+\frac{eV}{2}} T_{\sigma\sigma^\prime}(E)\,dE
	\label{equ7}
\end{equation}
where $e$ is the electronic charge and $h$ is the Plank constant. $T_{\sigma\sigma^\prime}(E)$ is the spin dependent two-terminal 
transmission probability. There are two possibilities of spin transmission. When $\sigma=\sigma^\prime$, we get pure spin transmission
($T_{\uparrow\uparrow}$, $T_{\downarrow\downarrow}$) and for $\sigma \ne \sigma^\prime$, spin-flip transmission occurs
($T_{\downarrow\uparrow}$, $T_{\uparrow\downarrow}$).
The net up and down spin transmission probabilities are defined as
\begin{eqnarray}
	\nonumber
	T_{\uparrow}=T_{\uparrow\uparrow}+T_{\downarrow\uparrow} \\
	\nonumber
	T_{\downarrow}=T_{\downarrow\downarrow}+T_{\uparrow\downarrow}.
\end{eqnarray}

Utilizing the Green's function formalism, we calculate the transmission probability for an electron injected with spin $\sigma$
and transmitted as $\sigma^\prime$, and it is expressed as~\cite{gf1,gf2}
\begin{equation}
	T_{\sigma\sigma^\prime} = \mbox{Tr}\left[\mbox{\boldmath{$\Gamma$}}_S^\sigma \mbox{\boldmath{$G$}}^r 
	\mbox{\boldmath{$\Gamma$}}_D^{\sigma^\prime} \mbox{\boldmath{$G$}}^a\right]
	\label{equ10}
\end{equation}
where $\mbox{\boldmath{$\Gamma$}}_{S(D)}^{\sigma(\sigma^\prime)}=-2 \mbox{Im}
\left[\mbox{\boldmath{$\Sigma$}}_{S(D)}^{\sigma(\sigma^\prime)}\right]$.
The matrices $\boldsymbol{\Sigma}_S^\sigma$ and $\boldsymbol{\Sigma}_D^{\sigma^\prime}$ are the self-energy corrections for the source 
and drain, respectively. The quantities $\mbox{\boldmath{$G$}}^r$ and $\mbox{\boldmath{$G$}}^a$ are the retarded and advanced Green's 
functions, respectively, and they can be calculated from the relation~\cite{gf1}
\begin{equation}
\mbox{\boldmath{$G$}}^r=(\mbox{\boldmath{$G$}}^a)^{\dagger}=\left[E \mbox{\boldmath{$I$}} - H_{\mbox{\tiny mol}} 
- \boldsymbol{\Sigma}_S^\sigma - \boldsymbol{\Sigma}_D^\sigma\right]^{-1}
\end{equation}
where $\mbox{\boldmath{$I$}}$ is the identity matrix of dimension $2N \times2N$. $N$ denotes the total number of magnetic sites in the helix.

\section{Numerical results and discussion}

Based on the above theoretical formulation, now we present and analyze the results which include (i) transmission-energy spectra, 
(ii) current-voltage and polarization-voltage characteristics, (iii) effect of geometrical conformation of the helical system, and 
(iv) the response of spin polarization with bias voltage and Fermi energy. Before going to the detailed analysis let us first describe 
the values of the parameters which remain fixed throughout the calculations. All the energies are measured in electron-volt (eV). For 
the AFH, we choose $\epsilon_i=0$ and $t_1=1$. The AFH is coupled to the source and drain via the coupling strength $t_S=0.8$ and 
$t_D=0.8$ respectively. 
\begin{table}[ht]
\caption{Physical parameters of the right-handed SRH and LRH helices.}
\vskip 0.15cm
\begin{tabular}{|c|c|c|c|c|} \hline 
AFH & $R$ ($\mbox\AA$) & $\Delta z$ ($\mbox\AA$) & $\Delta\phi$ & $l_c$ ($\mbox\AA$)  \\ \hline
SRH & $7$ & $3.4$ & $\frac{\pi}{5}$ & $0.9$  \\ \hline
LRH & $2.5$ & $1.5$ & $\frac{5 \pi}{9}$ & $0.9$  \\ \hline 
\end{tabular}
\label{tab1}
\end{table}
The on-site energy $\epsilon_0$ and the nearest-neighbor hopping integral $t_0$ for the electrodes are fixed at zero and $4$ respectively. 
We take the system size $N=20$ and compute the results of right-handed helices. The values of the structural parameters for the SRH and 
LRH helices are given in Table~\ref{tab1}, unless mentioned specifically, which are taken from the standard data set available 
in literature~\cite{rps}.

To achieve spin polarization, it is essential to ensure a distinguishable difference in the transmission probabilities of the up- 
and down-spin energy channels. However, in a perfect antiferromagnetic (AF) system, the transmission spectra for the two spin channels 
are identical. As a result, the corresponding spin currents become equal, leading to zero net spin polarization. This behavior arises 
because the full Hamiltonian of the system can be decoupled into two sub-Hamiltonians for the up- and down-spin electrons, each having 
an identical set of eigenvalues for their respective eigenstates. In our system, we introduce disorder in the magnetic moments to break 
this symmetry between the two sub-Hamiltonians. As a consequence, a finite mismatch in their eigenvalues emerges, which is essential for
obtaining non-zero spin polarization. We consider both correlated and uncorrelated disorders for two different AF configurations, and 
the spin filtration efficiency is critically examined for all these cases.

\subsection{Setup-1: with correlated disorder}

\begin{figure}[ht]
{\centering \resizebox*{8.5cm}{8.5cm}{\includegraphics{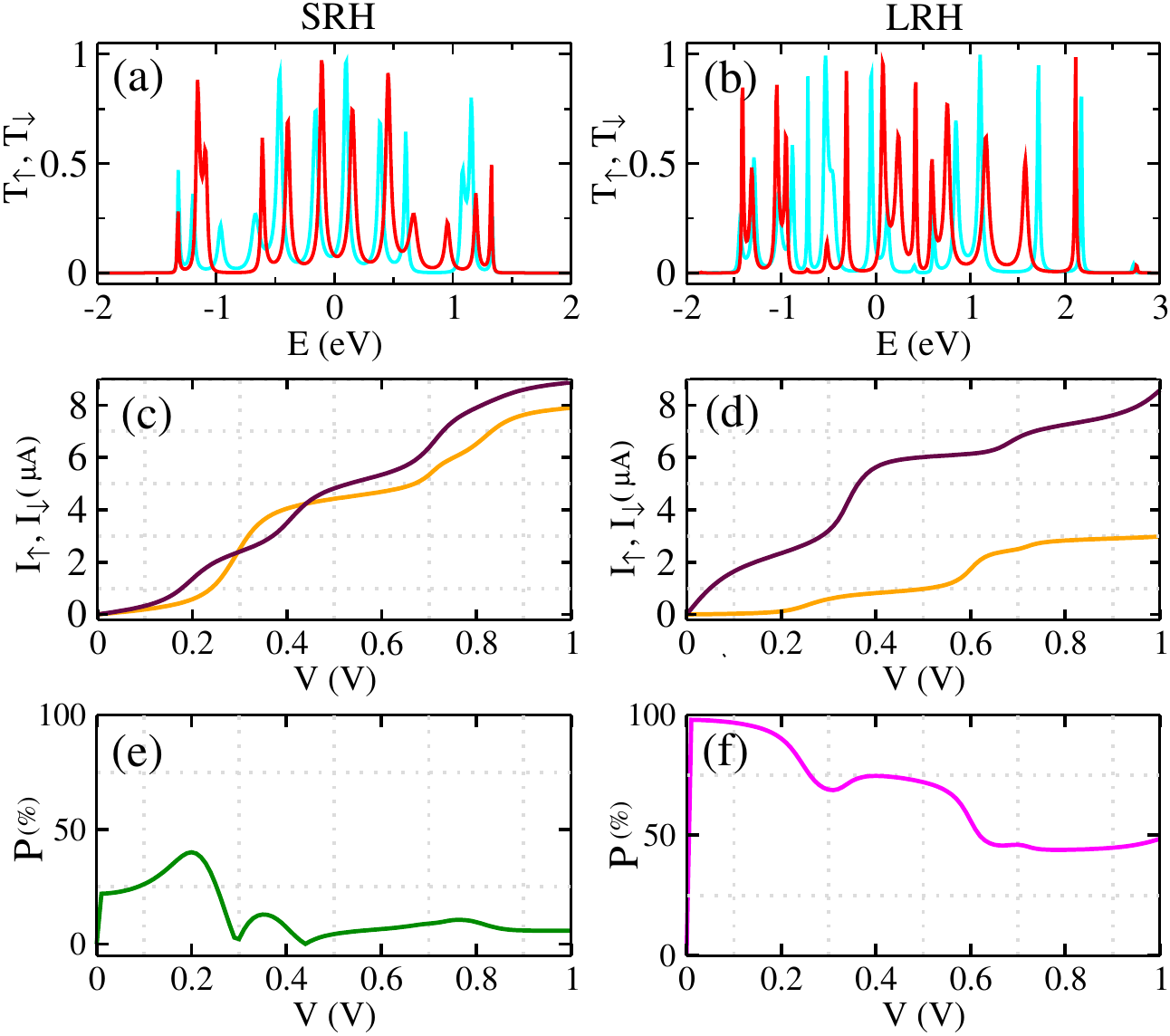}}\par}
\caption{(Color online). (a) and (b): Spin-dependent transmission probabilities as a function of energy where the cyan and red colors
correspond to the up and down spin electrons, respectively. (c) and (d): Variation of up and down spin currents with bias voltage where 
the orange and maroon colors represent the up and down spin currents, respectively. (e) and (f): Dependence of spin polarization on bias 
voltage for the SRH AFH (left column) and LRH AFH (right column) systems. Here, we fix Fermi energy at $E_F=0.25$.}
\label{fig2}
\end{figure}
\begin{figure}[ht]
{\centering \resizebox*{7.75cm}{7cm}{\includegraphics{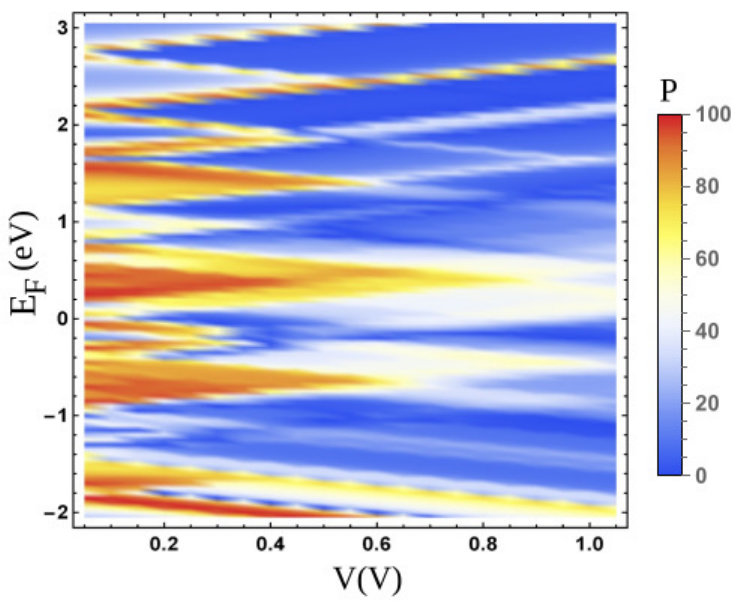}}\par}
\caption{(Color online). Simultaneous variation of spin polarization $P$ with Fermi energy $E_F$ and bias voltage $V$ for the LRH AFH.}
\label{fig3}
\end{figure}
\begin{figure}[ht]
{\centering \resizebox*{8.25cm}{4cm}{\includegraphics{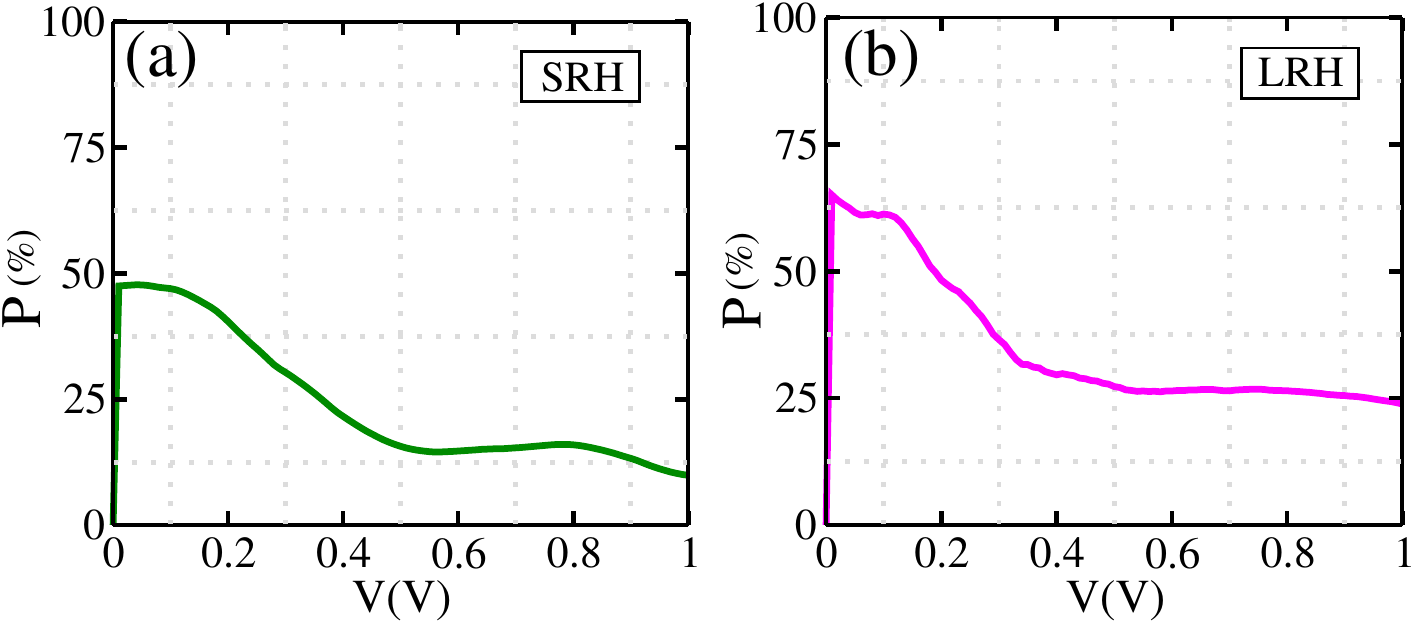}}\par}
\caption{(Color online). Dependence of spin polarization $P$ with bias voltage by fixing Fermi energy at $0.25$ for (a) SRH and (b) LRH 
helices.}
\label{fig4}
\end{figure}
\begin{figure}[ht]
{\centering \resizebox*{7.75cm}{7cm}{\includegraphics{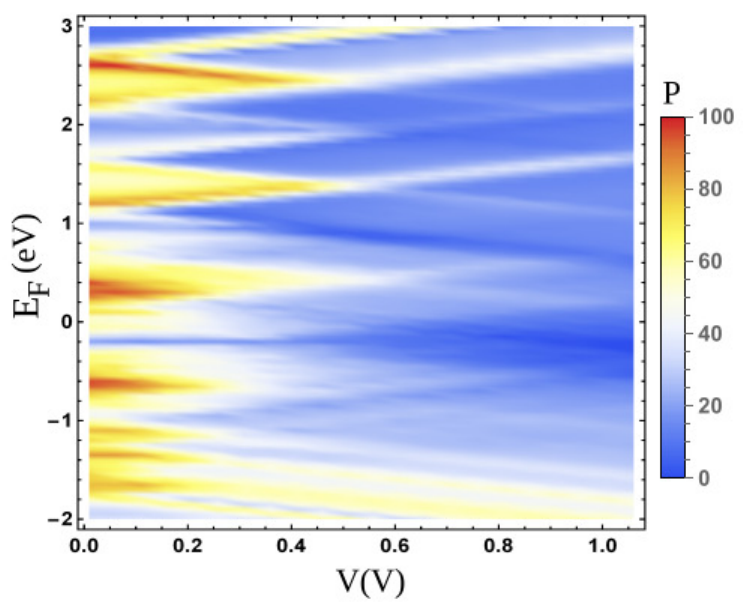}}\par}
\caption{(Color online). Variation of spin polarization with bias voltage and Fermi energy for the LRH AFH.}
\label{fig5}
\end{figure}

Figure~\ref{fig2} presents the results for the configuration shown in Fig.~\ref{model}(a), considering correlated disorder in the 
magnetic moments. To have a clear understanding, we begin by examining the transmission spectra of the nanojunction. Figures~\ref{fig2}(a) 
and \ref{fig2}(b) show the transmission probabilities for up-spin (cyan) and down-spin (red) electrons for AF helices with short-range 
hopping (SRH) and long-range hopping (LRH), respectively. A key observation is the finite mismatch between the transmission spectra of 
the two spin channels in both SRH and LRH helices. This arises due to the presence of disorder in the magnetic moments, which leads to 
distinct eigenvalues for the up- and down-spin sub-Hamiltonians. The separation between the up- and down-spin transmission peaks is more
pronounced in the LRH helix than in the SRH case. This can be attributed to the greater asymmetry of the LRH transmission spectrum around 
the band center, whereas the SRH spectrum remains nearly symmetric. A perfectly symmetric spectrum is recovered for the nearest-neighbor 
hopping limit. In the LRH case, within the energy range $-1 \le E \le 0$, the up-spin transmission dominates, while in the range 
$0 \le E \le 1$, the down-spin transmission becomes predominant. Consequently, by appropriately tuning the Fermi energy, one can 
achieve significant spin polarization.

Figures~\ref{fig2}(c) and \ref{fig2}(d) display the up-spin (orange) and down-spin (maroon) currents as a function of bias voltage for 
SRH and LRH helices, respectively, with the Fermi energy fixed at $E_F = 0.25\,$eV. Since the current is obtained by integrating the
transmission function, the features of the transmission spectra are directly reflected in the current-voltage characteristics. For the 
SRH case, the up- and down-spin currents remain nearly identical due to the small difference between their transmission probabilities. 
In contrast, for the LRH helix, electron transmission near the Fermi energy is dominated by the down-spin channel. As a result, when 
\begin{figure}[ht]
{\centering \resizebox*{8.25cm}{4cm}{\includegraphics{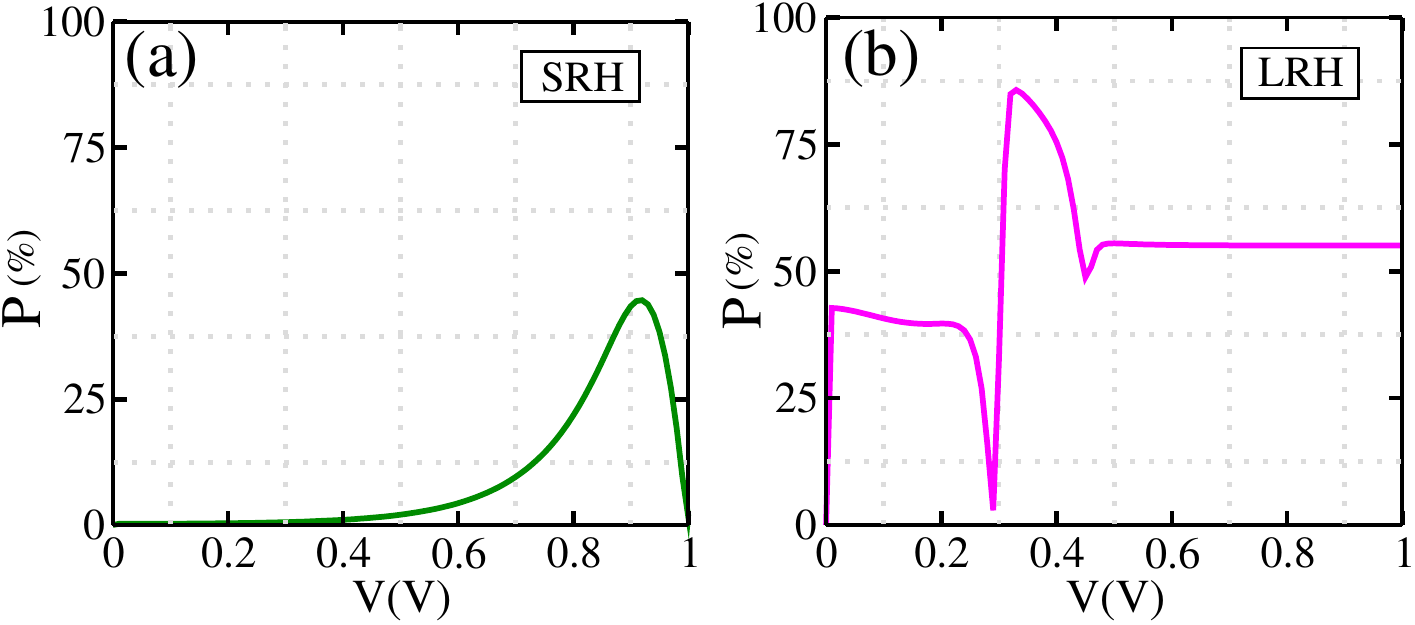}}\par}
\caption{(Color online). Variation of $P$ as a function of bias voltage $V$ at $E_F=0.1$, for (a) SRH and (b) LRH AF helices.}
\label{fig6}
\end{figure}
the Fermi energy lies in this region, a substantial down-spin current is obtained, while the up-spin current remains significantly 
suppressed. For bias voltages in the range $0 \le V \le 0.2$, the up-spin current is absent, whereas the down-spin current remains 
finite, yielding $\sim 100\%$ spin polarization in this voltage window. As the bias voltage increases, the difference between up- and 
down-spin currents becomes even more pronounced. Hence, in the LRH case, a high degree of spin polarization can be sustained even at 
large bias voltages. 

This behavior is clearly reflected in the polarization-voltage characteristics shown in Figs.~\ref{fig2}(e) and \ref{fig2}(f) for the 
SRH and LRH helices, respectively. For SRH, the spin polarization remains very small and approaches zero as the bias voltage increases. 
In contrast, the LRH case exhibits $\sim 100\%$ spin polarization at low bias, and a considerable degree of polarization persists even 
at higher voltages. 

Since the spin polarization efficiency depends on both the choice of Fermi energy and the applied bias voltage, it is important to 
inspect the influence of these parameters by varying them over a broad range. As the LRH helix exhibits superior spin filtration 
performance compared to its SRH counterpart, in Fig.~\ref{fig3} we present the variation of spin polarization as a function of Fermi 
energy and bias voltage for the LRH helix. Notably, multiple regions appear in the parameter space where the spin polarization reaches 
nearly $100\%$.

\subsection{Setup-1: with uncorrelated disorder}

From the above analysis, we find that for the antiferromagnetic configuration of setup-1, a high degree of spin polarization is achieved
\begin{figure}[ht]
	{\centering \resizebox*{7.75cm}{7cm}{\includegraphics{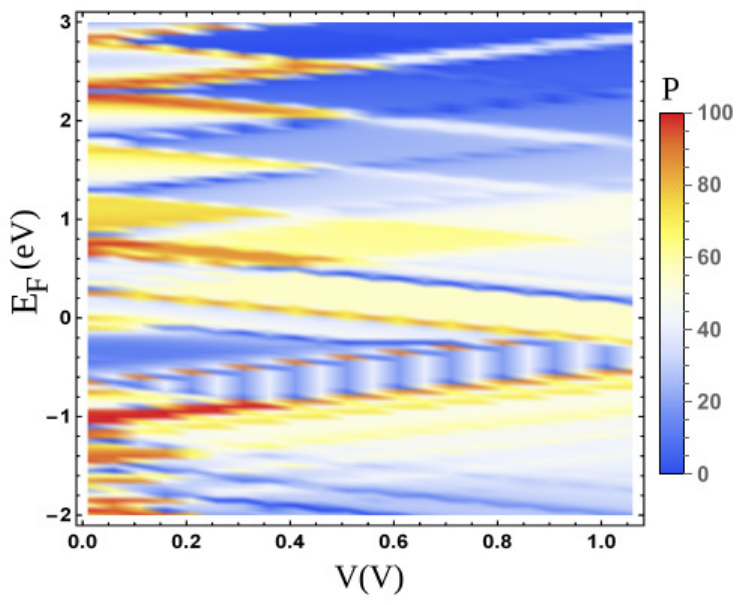}}\par}
	\caption{(Color online). Simultaneous variation of $P$ with bias voltage and Fermi energy for the LRH AFH.}
	\label{fig7}
\end{figure}
in the LRH case when correlated disorder in the magnetic moments is considered. To make the analysis more general, we now investigate the 
same antiferromagnetic configuration with uncorrelated (random) disorder in the magnetic moments. Figure~\ref{fig4} shows the variation of 
spin polarization as a function of bias voltage for both SRH and LRH helices, with the Fermi energy fixed at $E_F=0.25$ eV. For the LRH 
helix, a spin polarization of about $70\%$ is obtained at low bias, which gradually decreases as the bias voltage increases. In contrast, 
the maximum spin polarization for the SRH case is approximately $50\%$. At higher bias voltages, a larger number of up- and down-spin 
channels contribute to the current, reducing the difference between the two spin currents and thereby lowering the overall spin 
polarization. Thus, for random disorder in the magnetic moments, a substantial degree of spin polarization is obtained for the LRH helix 
at low bias. 

We further examine the spin filtration efficiency across all relevant values of Fermi energy and bias voltage. Figure~\ref{fig5} illustrates 
the variation of spin polarization with simultaneous changes in both parameters for the LRH helix. We observe that a high level of spin
polarization is achieved for multiple values of the Fermi energy. Therefore, by appropriately selecting the Fermi energy and bias voltage, 
one can obtain a significant amount of spin polarization.

\subsection{Setup-2: with correlated disorder}

\begin{figure}[ht]
{\centering \resizebox*{8.5cm}{4cm}{\includegraphics{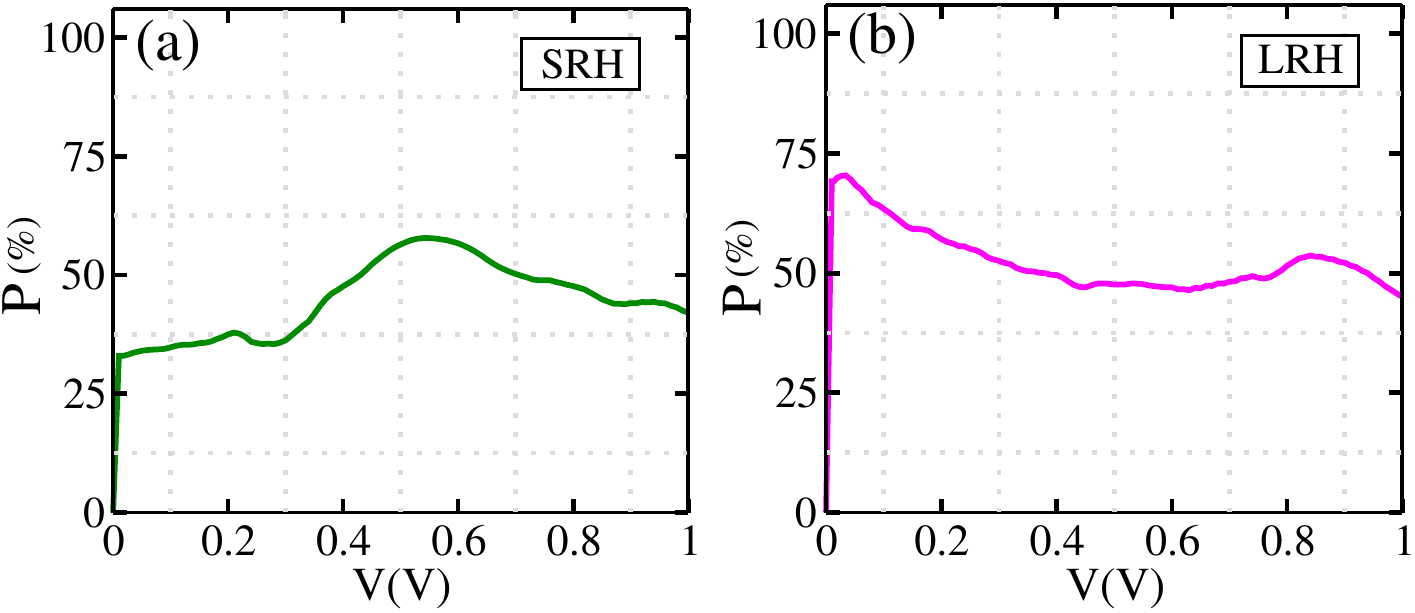}}\par}
\caption{(Color online). Polarization-voltage characteristics for (a) SRH and (b) LRH helices, when the Fermi energy is set at $0.3$.}
\label{fig8}
\end{figure}
\begin{figure}[ht]
{\centering \resizebox*{7.75cm}{7cm}{\includegraphics{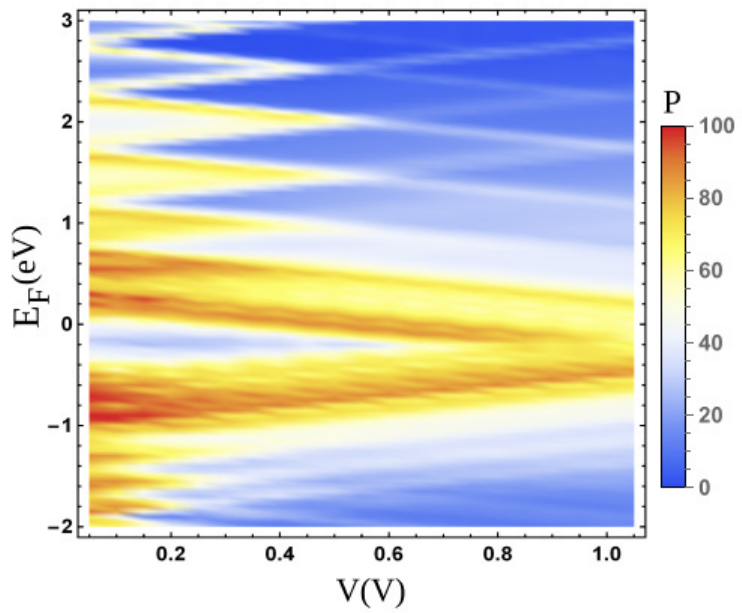}}\par}
\caption{(Color online). Simultaneous variation of $P$ with Fermi energy and bias voltage for LRH AFH.}
\label{fig9}
\end{figure}
\begin{figure*}[ht]
{\centering \resizebox*{14cm}{12cm}{\includegraphics{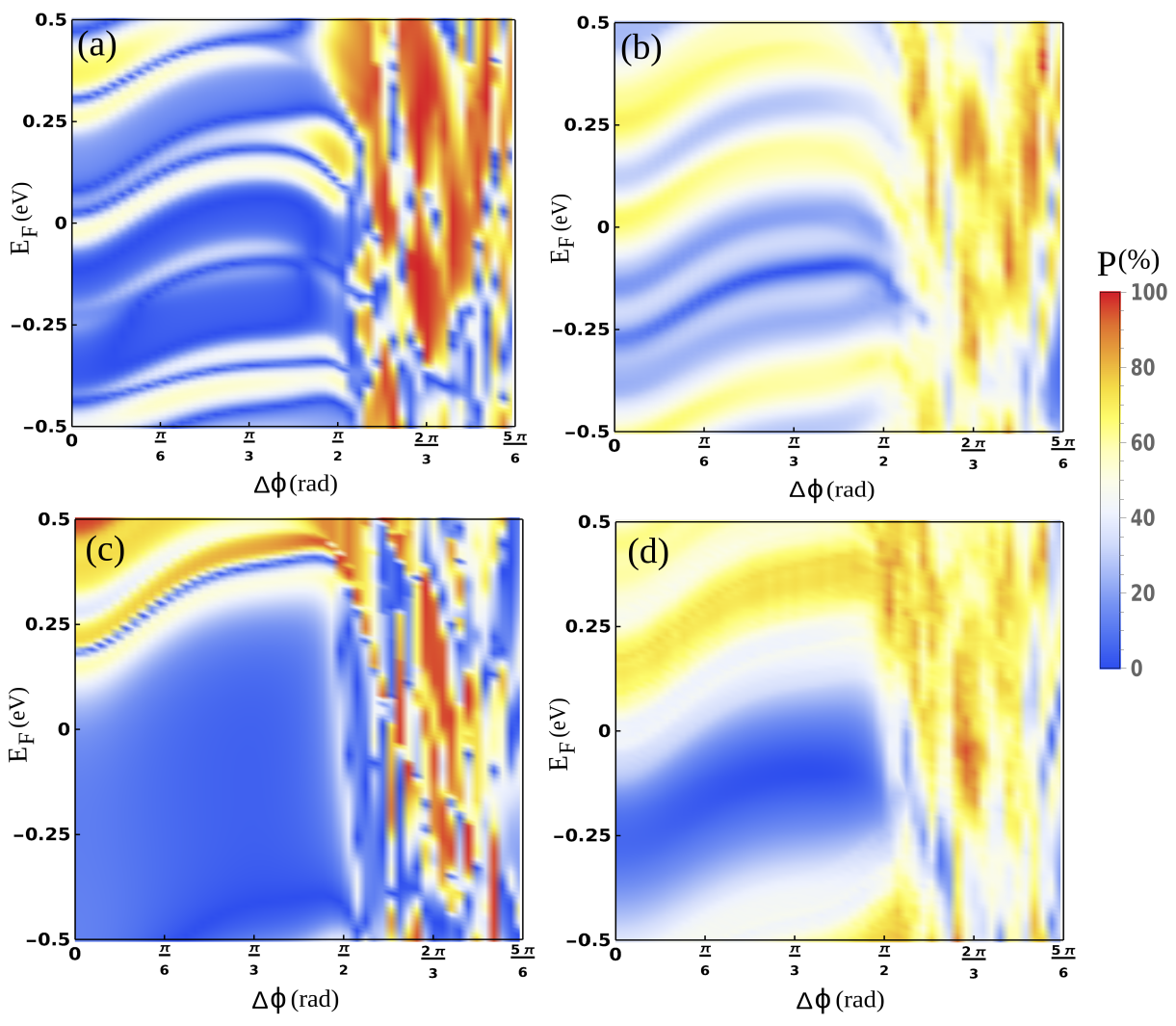}}\par}
\caption{(Color online). Simultaneous variation of $P$ with Fermi energy and twisting angle, for (a) setup-1 with correlated disorder, 
(b) setup-1 with random disorder, (c) setup-2 with correlated disorder, and (d) setup-2 with random disorder. The other parameters are:
$V=0.2$, $R=2.5$, $\Delta z=1.5$, and $l_c=0.9$.}
\label{fig10}
\end{figure*}

In this sub-section, we consider a different antiferromagnetic ordering, as shown in Fig.~\ref{model}(b), and introduce disorder in the 
magnetic moments following the AAH model. For both SRH and LRH helices, we compute the spin polarization efficiency as a function of bias
voltage by fixing the Fermi energy. Figure~\ref{fig6} presents the $P$-$V$ characteristics at $E_F=0.1$. Here too, the LRH helix exhibits 
a favorable response, showing a high degree of spin polarization even at moderate bias voltages. This enhanced performance arises from the
combined effects of disorder and higher-order hopping, which significantly alter the up- and down-spin channels. Due to long-range hopping, 
the energy spectrum becomes more asymmetric and develops a gap-like structure in the presence of correlated disorder. Consequently, a
substantial mismatch emerges between the up- and down-spin currents, resulting in a high degree of spin polarization. In contrast, for the 
SRH case, the spin polarization remains small. Without long-range hopping, the spectrum is less asymmetric, allowing both spin channels 
to contribute to the current, which reduces the overall polarization.

For completeness, we also examine the dependence of spin polarization on a broad range of Fermi energies and bias voltages for the LRH 
helix. The results, shown in Fig.~\ref{fig7}, indicate that a significant level of spin polarization can be achieved across various 
bias voltages for different choices of the Fermi energy.

\subsection{Setup-2: with uncorrelated disorder}

We now explore how the spin filtration efficiency is affected when random disorder is introduced in the magnetic moments of setup-2, 
shown in Fig.~\ref{model}(b). Figure~\ref{fig8} illustrates the variation of spin polarization as a function of bias voltage for both SRH 
and LRH helices, with the Fermi energy fixed at $E_F = 0.3$. A reasonable amount of spin polarization is obtained for both helices over 
a wide range of bias voltages, however, the LRH case exhibits a more favorable response.

A clearer understanding emerges from Fig.~\ref{fig9}, where the spin polarization is plotted as a function of both Fermi energy and bias
voltage. Within the range $-1 \le E_F \le 1$, several Fermi energy values yield significant spin polarization. Moreover, a considerable 
amount of polarization persists even under high bias voltages.

\subsection{Conformational effect} 

To obtain a comprehensive understanding of how geometrical conformation influences spin polarization, we calculate the spin polarization 
for different values of $\Delta \phi$ and Fermi energy across all four possible cases. The results are presented in Fig.~\ref{fig10}, 
where sub-figures (a), (b), (c), and (d) correspond to setup-1 with correlated disorder, setup-1 with random disorder, setup-2 with 
correlated disorder, and setup-2 with random disorder, respectively. The twisting angle is varied from $0$ to $5\pi/9$, while the Fermi 
energy ranges from $-0.5$ to $0.5$. Across all scenarios, we observe that when the twisting angle is zero, indicating the absence of
helicity, the spin polarization remains very low for most values of the Fermi energy. In contrast, for twisting angles between $\pi/2$ 
and $5\pi/6$, significant spin polarization is achieved over a broad range of Fermi energies. These observations clearly highlight the 
crucial role of helicity in generating substantial spin polarization.

\section{Closing remarks}

To conclude, in the present work, we have investigated spin-dependent transport in an antiferromagnetic helix with a nonuniform 
distribution of magnetic moments. Two distinct antiferromagnetic configurations have been considered, and to ensure generality, both 
correlated and uncorrelated disorders have been incorporated. The interplay between short-range hopping and long-range hopping has 
been examined in detail. The nanojunction, comprising the AFH clamped between source and drain electrodes, has been modeled within 
a tight-binding framework. Transmission probabilities have been evaluated using the Green's function formalism, while spin-dependent 
currents have been computed following the Landauer-B\"uttiker prescription. The spin polarization efficiency has then been extracted 
from these currents.

For all four configurations (setup-1 with correlated disorder, setup-1 with random disorder, setup-2 with correlated disorder, and 
setup-2 with random disorder), we have systematically analyzed the spin filtration efficiency under various input conditions. To 
illustrate the underlying mechanisms more clearly, we have presented the spin-resolved transmission spectra and current-voltage 
characteristics for one representative case. The key findings are summarized below:

\vskip 0.2cm
\noindent
$\bullet$ Introducing disorder in the magnetic moments breaks the symmetry between the spin channels. As a result, the up- and 
down-spin transmission probabilities become distinct, leading to finite spin polarization even in the absence of an external electric field.\\
$\bullet$ The current-voltage characteristics reveal that, for the LRH AFH, a substantial separation between up- and down-spin currents 
persists over a wide range of bias voltages, yielding a high degree of spin polarization. \\
$\bullet$ Across all four configurations, large spin polarization values are obtained for broad parameter ranges, demonstrating the 
robustness of our results.\\
$\bullet$ A comparison between SRH and LRH helices clearly indicates that the LRH configuration consistently provides a more favorable 
spin filtration response.

Our findings suggest that antiferromagnetic helices with non-uniform magnetic moments and long-range hopping offer a promising platform 
for designing efficient spintronic devices.

\section*{ACKNOWLEDGMENT}

SS is thankful to ANRF, India (File number: PDF/2023/000319) for providing her research fellowship.

\end{document}